\documentclass[twocolumn,showpacs,preprintnumbers,amsmath,amssymb,widetext]{revtex4}
\usepackage{graphicx}
\begin{document}
\title {Cold electron Josephson transistor}

\author{A.M. Savin}
\author{J.P. Pekola}
\author{J.T. Flyktman}
\author{A. Anthore}
\affiliation{Low Temperature Laboratory, Helsinki University of Technology, P.O. Box 2200, FIN-02015 HUT, %%@
Finland}
\author{F. Giazotto}
\affiliation{NEST-INFM \& Scuola Normale Superiore, I-56126 Pisa,
Italy}

\begin{abstract}
A superconductor-normal metal-superconductor mesoscopic Josephson junction has been realized in which %%@
the critical current is tuned through normal current injection using a symmetric electron cooler directly connected %%@
to the weak link.
Both enhancement of the critical current by more than a factor of two, and supercurrent suppression have been %%@
achieved by varying the cooler bias. Furthermore, this
transistor-like device demonstrates large current gain ($\sim$20)
and low power dissipation.

\end{abstract}


\maketitle

Transport dynamics in mesoscopic structures where normal metals (N) are coupled with  superconductors (S) are %%@
nowadays in the focus of extensive research \cite{belzig,lambert}.
This stems mainly from the relevance these systems have both from the fundamental physics point of view  and %%@
in light of their possible exploitation in nanoelectronics. In diffusive SNS junctions, where the length of the N region exceeds the elastic mean free path, coherent %%@
sequential Andreev scattering \cite{andr} between the superconductors may lead to a continuum spectrum of resonant levels %%@
\cite{belzig} responsible for carrying the supercurrent flow through the  structure. The %%@
Josephson current  is given by supercurrent spectrum  weighted by the occupation number of %%@
correlated electron-hole pairs that is  determined by the quasiparticle energy distribution in the N region %%@
of the junction. By changing the latter through current injection from additional \textit{nonsuperconducting} %%@
terminals connected to the N region \cite{wilhelm} both supercurrent suppression \cite{morpurgo} as well as %%@
its sign reversal ($\pi$-transition) were demonstrated
\cite{baselmans}. As  predicted in Refs. \cite{giazotto1, baselmans3}, the distinctive quasiparticle %%@
distribution existing in the N region of a biased SINIS structure (where I stands for an insulating barrier) %%@
is also well suited to control the Josephson coupling in a long SNS weak link, allowing indeed either large supercurrent %%@
\textit{enhancement} or efficient suppression with respect to equilibrium. %%@

\begin{figure}[b!]
\begin{center}
\includegraphics[width=7.9cm,clip]{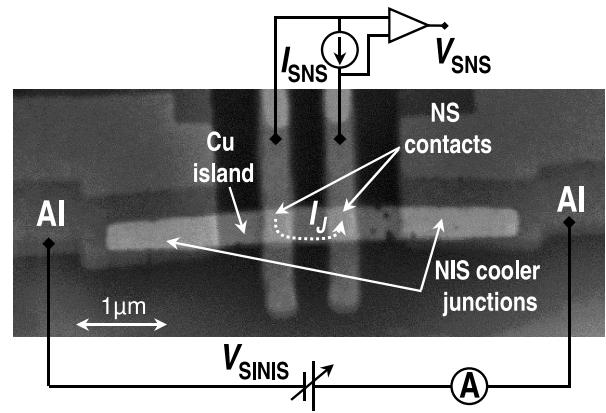}
\end{center}
\caption{Scanning electron micrograph of a typical structure including a sketch of the measurement circuit. %%@
Two superconducting Al  electrodes are connected through insulating barriers to a Cu island so to realize %%@
a symmetric SINIS electron cooler. The supercurrent $I_J$ in the Al/Cu/Al junction is tuned upon voltage %%@
biasing the SINIS control line.} \label{device}
\end{figure}

In this Letter, we present the implementation and  characterization of a  %%@
four-terminal superconducting structure (see Fig. \ref{device})  consisting of a SNS mesoscopic junction %%@
integrated with a SINIS electron cooler. A similar device was considered but not successfully operated %%@
in Ref. \cite{baselmans3}.  In this transistor, the maximum supercurrent flowing in the SNS junction is %%@
controlled by voltage biasing the SINIS line whose N region is
shared with the Josephson junction.
Low temperature transport measurements show enhancement of the critical current %%@
under hot quasiparticle extraction by more than a factor of two  with respect to equilibrium. In %%@
addition this device demonstrates low power dissipation and large
current gain.

\begin{figure}[t!]
\begin{center}
\includegraphics[width=8.0cm,clip]{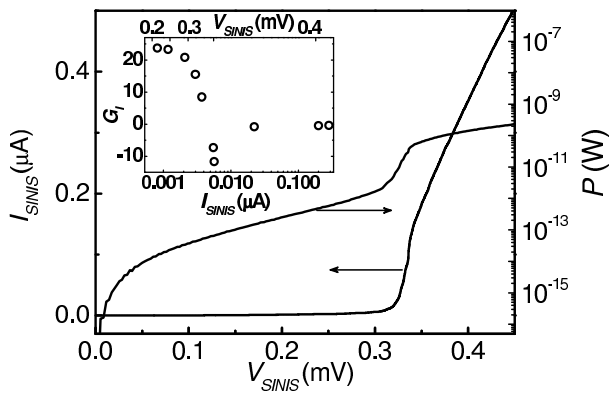}
\end{center}
\caption{Current-voltage characteristic (left axis) and power dissipation $P=V_{SINIS}I_{SINIS}$ (right axis) %%@
of the SINIS line at $T_{bath}=72$ mK. The inset shows the measured differential current gain $\mathbf{G}_I$ %%@
against  $I_{SINIS}$ at the same temperature (also displayed is
the dependence on $V_{SINIS}$).} \label{performance}
\end{figure}

The sample (shown in Fig. \ref{device}) consists of a Cu island, $0.37\,\mu$m wide and 30 nm thick, %%@
symmetrically connected  at its ends via insulating barriers  (with normal-state resistance $\mathcal{R}_T %%@
\simeq 240\,\Omega$) to two 60-nm-thick Al reservoirs, thus realizing a SINIS cooler. %%@
The Josephson junction instead consists of an Al/Cu/Al SNS weak link (with normal-state resistance %%@
$\mathcal{R}_N =11.5\,\Omega$), whose N region is shared with the SINIS line. The  minimum interelectrode %%@
separation in the SNS junction of the present device is $L_J \simeq 0.4\,\mu$m.
The structure was fabricated on a thermally oxidized Si  substrate by electron beam lithography and %%@
three-angle shadow-mask evaporation. The electrical characterization was performed at different bath %%@
temperatures down to 70 mK in a dilution refrigerator. From  low-temperature  resistance measurements %%@
we deduced the Cu diffusion coefficient $D \approx 10$ cm$^2$/s.
This low value of $D$ is probably caused by significant
intermixing of the materials at the NS interface leading to the
strong reduction of the
electron mean free path in the weak link. The Al energy gap, $\Delta=169\,\mu e$V, %%@
was inferred from the low-temperature current-voltage %%@
characteristic of the SINIS  line (see Fig. \ref{performance}). The %%@
coherence length $\xi _N=\sqrt{\hbar D/\Delta}\approx 62$ nm is
then much smaller than $L_J$, providing the frame of the
\textit{long} junction regime.

\begin{figure}[b!]
\begin{center}
\includegraphics[width=8.5cm,clip]{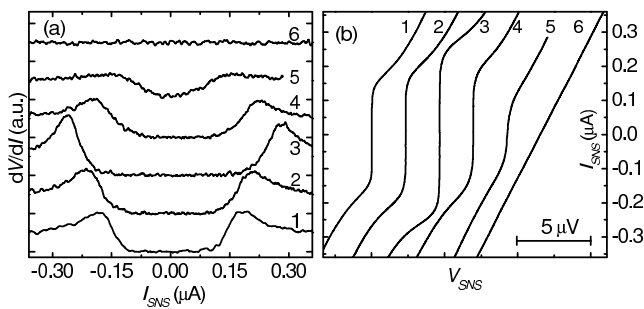}
\end{center}
\caption{Selected $dV/dI$ vs $I_{SNS}$ (a) and current-voltage characteristics (b) of the SNS junction at  %%@
$T_{bath}=72$ mK for different $V_{SINIS}$ values (all curves are offset for clarity): 1 - 0, 2 - 194 $\mu$V, 3 - %%@
300 $\mu$V, 4 - 342 $\mu$V, 5 - 355 $\mu$V, 6 - 938 $\mu$V.  Curves in (b) were obtained by numerical %%@
integration of the corresponding ones in (a).}
\label{current/voltage}
\end{figure}

The experiment consists of sweeping the $I_{SNS}$ current across the SNS junction while measuring its %%@
differential resistance $dV/dI$ at different values of voltage bias   %%@
($V_{SINIS}$) across the SINIS control line. Figure %%@
\ref{current/voltage}(a) shows a subset of $dV/dI$ vs $I_{SNS}$ characteristics measured at the bath %%@
temperature $T_{bath}=72$ mK for several $V_{SINIS}$. The curves display a nonhysteretic behavior %%@
characteristic for overdamped junctions \cite{likharev}. In the
case of an SNS weak link the effect of thermal fluctuations on the
smearing of the voltage-current characteristic is stronger
\cite{hoss} than predicted by the model for resistively shunted
junction \cite{ambeg}. We have chosen to define the experimental critical current as the current where the differential resistance reaches %%@
$\mathcal{R}_N/2$ \cite{dubos}. Notably, upon increasing $V_{SINIS}$,  the current range where the %%@
differential resistance vanishes widens initially, thus reflecting an enhancement of $I_{J}$, being maximized %%@
at a voltage  corresponding to $V_{SINIS}=300$ $\mu$V $ \simeq 1.8\,\Delta /e$ \cite{leivo} (curve labelled as 3 in %%@
Fig. \ref{current/voltage}(a)); then, further increase of bias leads to a monotonic decay and to a complete %%@
suppression of $I_{J}$ at larger voltages (curve labelled as 6 in
Fig. \ref{current/voltage}(a)). This non-monotonic behavior is seen in the corresponding $I-V$ curves  in Fig. \ref{current/voltage}(b). %%@

\begin{figure}[t!]
\begin{center}
\includegraphics[width=8.0cm,clip]{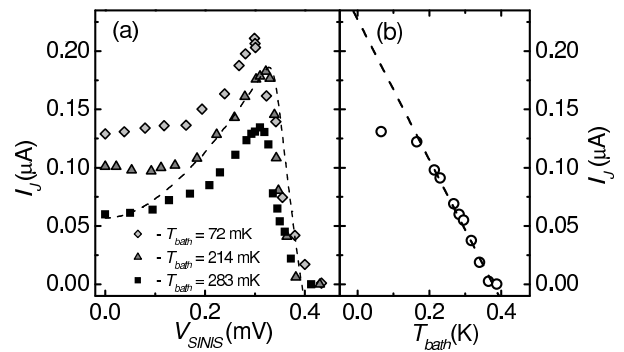}
\end{center}
\caption{(a) Critical current $I_J$ vs control voltage $V_{SINIS}$ at three different bath temperatures. %%@
(b) Equilibrium supercurrent ($V_{SINIS}=0$) vs bath temperature. Dashed %%@
line in (a) represents curve obtained from energy balance equation (\ref{baleq}) and the linear approximation  %%@
of  $I_J (T_{bath})$ shown in (b).} \label{supercurrentbehavior}
\end{figure}

The observed behavior is due to the  relation existing between the %%@
observable maximum supercurrent $I_J$ and the quasiparticle energy
distribution  in the weak link. In the present experimental situation of large $L_{SINIS}$, inelastic %%@
electron-electron relaxation forces  the electron system to retain a local %%@
thermal (quasi)equilibrium. As a consequence,  the quasiparticle energy distribution can be described with a %%@
Fermi-Dirac function at an \textit{effective} electron temperature
$T_e$. The temperature $T_e$ is determined by the balance between
two heat flows :
\begin{equation}
\mathcal{P}(V_{SINIS},T_e,T_{bath})+\mathcal{P}_{e-bath}(T_e,T_{bath})=0.
\label{baleq}
\end{equation}
The first term accounts for the net heat current %%@
$\mathcal{P}$ transferred from the N island to the superconductors
upon biasing the SINIS line \cite{leivo}:
\begin{equation}
\mathcal{P}=\frac{2}{e^2\mathcal{R}_T}\int_{-\infty}^\infty %%@
n(E)[f_0(\tilde{E},T_e)-f_0(E,T_{bath})]\tilde{E}dE,
\label{coolingpower}
\end{equation}
where $\tilde{E}=E-eV_{SINIS}/2$, $f_0(E,T)$ is the Fermi-Dirac distribution function  and %%@
$n(E)=|\text{Re}[(E+i\Gamma)/\sqrt{(E+i\Gamma)^2-\Delta^2}]|$
is the (smeared by nonzero $\Gamma$) BCS density of states of the superconductor \cite{gamma}.
Equation (\ref{coolingpower}) is symmetric in $V_{SINIS}$, being maximized slightly below $|2\Delta /e|$. %%@
The second term accounts for energy transfer from electrons to the phonons of the normal island at the temperature $T_{bath}$ %%@
and is equal to %%@
$\mathcal{P}_{e-bath}=\Sigma \mathcal{V}(T^5_e -T^5_{bath})$ \cite{urbina}, where $\mathcal{V}$ %%@
is the volume of the N island and $\Sigma \approx 2$ nWK$^{-5}\mu$m$^{-3}$ for copper \cite{leivo}. The temperature %%@
$T_e$ in the weak link thus strongly depends on $V_{SINIS}$ and
can be smaller than $T_{bath}$ \cite{roukes}. At low temperature
(i.e., $k_B T_{bath}\ll \Delta $), in a long SNS junction, $I_J$
is predicted to decrease exponentially as $T_e$ increases
\cite{sup} in the regime where $k_B T_e \gg E_{Th} =\hbar D/L_J^2$
. Thus, upon biasing the SINIS line, $I_J$ will be changed with
respect to equilibrium (i.e., at $V_{SINIS}=0$), due to the
modification of $T_e$ that now differs from $T_{bath}$.

In Fig. \ref{supercurrentbehavior}(a) we plot the extracted $I_{J}$ values as a function of $V_{SINIS}$ at three different bath %%@
temperatures. For all displayed temperatures, the critical current increases monotonically up to about %%@
$V_{SINIS}\simeq 1.8\,\Delta /e$ as expected from the reduction of $T_e$ by cooling. Then, further increase of bias voltage leads to an efficient %%@
supercurrent suppression due to electron heating. %%@
The equilibrium critical current (i.e., at $V_{SINIS}=0$) vs  %%@
$T_{bath}$ is displayed in Fig. \ref{supercurrentbehavior}(b). The $I_J$ behavior follows a %%@
characteristic trend, decreasing upon rising the temperature, but it differs from the temperature dependence predicted %%@
by quasiclassical Green-function theory \cite{belzig}. The discrepancy can be ascribed to the uncertainty in the determination of the actual values of %%@
critical current, relatively narrow temperature range where it was
observed and thermal decoupling between electrons and bath at
temperatures below 200 mK. In Fig. \ref{supercurrentbehavior}, we show the expected critical current dependence on $V_{SINIS}$ %%@
at $T_{bath}=283$ mK obtained from the solution of Eqs. (\ref{coolingpower}) and (\ref{baleq}) to determine %%@
the effective electron temperature $T_e$ upon biasing the SINIS line, and assuming a linear behavior of the critical %%@
current $I_J$ versus $T_e$ below about 350 mK, the slope of the linear dependence being inferred from the measured $I_{J}(T_{bath})$. %%@
For this calculation, we assumed the already given parameters for the %%@
SINIS line and $\Gamma =1.8\cdot 10^{-3}\Delta$ estimated from the ratio ($\simeq %%@
\Gamma/\Delta$) of the low-temperature  SINIS conductance at low
and high bias \cite{pekola}. The resemblance between calculation
and experiment is evident although details of the former one are
dictated by the $I_J$ dependence on temperature, which we cannot
extrapolate reliably.
To better characterize our device, we show in Fig. \ref{performance} (right axis) the dissipated power $P$ %%@
against $V_{SINIS}$ in the SINIS line at $T_{bath}=72$ mK. The plot reveals that in the bias voltage region %%@
of significant critical current enhancement (i.e., in the $200 - 300\,\mu$V bias range) $P$ obtains values of %%@
the order of $10^{-13}$ W, while in the  regime of supercurrent suppression (i.e., for $V_{SINIS}>300\,\mu$V) %%@
some tens of pW. This demonstrates the low power dissipation intrinsic to the structure %%@
\cite{giazotto1}.
The $P$ behavior is directly related to the normal current flow in the control line. The latter is %%@
displayed in the left axis of Fig. \ref{performance} and shows that control currents as low as  a few %%@
nA are necessary to enhance  the critical current, while of about 100 nA to suppress it. %%@
The differential current gain $\mathbf{G}_I =dI_{J}/dI_{SINIS}$ against $I_{SINIS}$ is shown in the %%@
inset of Fig. \ref{performance}. Notably, $\mathbf{G}_I$ obtains values exceeding $20$ in the hot %%@
quasiparticle extraction regime, while of about $-11$ in the voltage region of supercurrent suppression. %%@
We  note that higher $\mathbf{G}_I$ values, as well as lower power dissipation and control %%@
currents, could be attained by optimizing the structure design
\cite{giazotto1, pekola}.

In summary, we have demonstrated experimentally control
of  Josephson coupling under hot quasiparticle extraction in a four-terminal superconducting structure. Our %%@
experimental result shows the potential of a SINIS line as a basis of a promissing class of mesoscopic %%@
transistors with high current gain. %%@

We acknowledge T.T. Heikkil\"{a}, F. Carillo, R. Fazio, P.J. Hakonen, F.W.J. Hekking and F. Taddei  for discussions, %%@
and Academy of Finland for financial support (TULE program). One of us (F. %%@
G.) would like to acknowledge the Large Scale Installation Program ULTI-3 of the European Union for the kind %%@
hospitality and for financial support.

%------------------------------------------ References

\end{document}